\documentclass[notitlepage,a4,10.5pt]{article}

\usepackage{graphicx}

\usepackage{amsmath}%
\usepackage{amsfonts}%
\usepackage{amssymb}%
\usepackage{mathrsfs}
\usepackage{amsthm}
\usepackage{float}

\usepackage{authblk}

\usepackage[left=2.5cm,right=2.5cm,top=3cm,bottom=3cm]{geometry}

\newcommand{\mc}{\mathcal}
\newcommand{\cp}{\times}

\newcommand{\Norm}[1]{\left\lvert\left\lvert #1 \right\rvert\right\rvert}

\newcommand{\bol}{\boldsymbol}

\newcommand{\abs}[1]{\left\lvert{#1}\right\rvert}

\newcommand{\lr}[1]{\left({#1}\right)}

\newcommand{\p}{\partial}

\newcommand{\ti}[1]{\textit{#1}}
\newcommand{\tb}[1]{\textbf{#1}}

\begin{document}

\title{Realization of Incompressible Navier-Stokes Flow as Superposition of Transport Processes for Clebsch Potentials}
\author[1]{Naoki Sato} 
\affil[1]{Graduate School of Frontier Sciences, \protect\\ The University of Tokyo, Kashiwa, Chiba 277-8561, Japan \protect\\ Email: sato\_naoki@edu.k.u-tokyo.ac.jp}
\date{\today}
\setcounter{Maxaffil}{0}
\renewcommand\Affilfont{\itshape\small}

    \maketitle
    \begin{abstract}
				
		In ideal fluids, Clebsch potentials occur as paired canonical variables associated with the Hamiltonian description of the Euler equations.
		This paper explores the properties of the incompressible Navier-Stokes equations when the velocity field is expressed through a complete set of paired Clebsch potentials.  
		First, it is shown that the incompressible Navier-Stokes equations can be cast as a system of transport (convection-diffusion) 
		equations where each Clebsch potential plays the role of a generalized distribution function. 
		The diffusion operator associated with each Clebsch potential departs from the standard Laplacian due to a term 
		depending on the Lie-bracket of the corresponding Clebsch pair. 
		It is further shown that the Clebsch potentials can be used to define a Shannon-type entropy measure, i.e. a functional, different from energy and enstrophy,  
		whose growth rate is non-negative. As a consequence, the flow must vanish at equilibrium. 
		This functional can be interpreted as a measure of the topological complexity of the velocity field.
		In addition, the Clebsch parametrization enables the identification of a class of flows, larger than the class of two dimensional flows, 
		possessing the property that the vortex stretching term identically vanishes and the growth rate of entrophy is non-positive.  
		\end{abstract}

\section{Introduction}
In this study we are concerned with the incompressible Navier-Stokes equations describing the motion of a viscous fluid in 
a bounded region $\Omega$ with boundary $\p\Omega$ contained in three dimensional Euclidean space $E^3$. Denoting with $\bol{x}=\lr{x,y,z}^T$ Cartesian coordinates in $E^3$, with $t$ the time variable, with $\bol{v}\lr{\bol{x},t}$ the velocity field, with $v_i\lr{\bol{x},t}$, $i=1,2,3$, the Cartesian components of $\bol{v}$, with $P\lr{\bol{x},t}$  the pressure field, and with $\nu$ a positive real constant (kinematic viscosity), the equations can be written as:     
\begin{subequations}
\begin{align}
\frac{\p\bol{v}}{\p t}=&\bol{v}\cp\lr{\nabla\cp\bol{v}}-\nabla\lr{\frac{\bol{v}^2}{2}+P}-\nu\nabla\cp\nabla\cp\bol{v},\label{NS1}\\
\nabla\cdot\bol{v}=&0.\label{NS2}
\end{align}
\end{subequations}\label{NS}
System \eqref{NS} is equipped with boundary conditions, 
\begin{equation}
\bol{v}=\bol{0}~~~~{\rm on}~~\p\Omega,\label{BC}
\end{equation}
and initial data $\bol{v}_0=\bol{v}\lr{\bol{x},0}$ at $t=0$. 

Our aim in this paper is to elucidate certain properties of the incompressible Navier-Stokes system
resulting from a Clebsch parametrization of the
velocity field \cite{Lin,Fukagawa},
\begin{equation}
\bol{v}=\nabla\phi+\sum_{i=1}^{N}p^i\nabla q^i,\label{Clebsch0}
\end{equation} 
where $\phi\lr{\bol{x},t}$, $p^i\lr{\bol{x},t}$, $q^i\lr{\bol{x},t}$, $i=1,...,N$, are Clebsch potentials and $N\geq 2$ is
a natural number. The representation \eqref{Clebsch0} is complete for $N\geq 2$, 
i.e. the number $2N+1$ of Clebsch potentials is sufficient to express an arbitrary vector field in three dimensional Euclidean space. 
Furthermore, when $N\geq 3$, there is enough freedom to enforce desired boundary conditions on the potentials themselves \cite{YosClebsch,YosDuality}.  

The Clebsch parametrization \eqref{Clebsch0} can be used to cast 
the incompressible Navier-Stokes equations as a system of Boltzmann-type
transport equations (see \cite{YosEpi2D,Scholle16} which treat the cases $\nu=0$ and $N=1$ respectively):
\begin{equation}
\frac{\p p^i}{\p t}+\nabla\cdot\lr{p^i\bol{v}}=\nu C^{p^i},~~~~\frac{\p q^i}{\p t}+\nabla\cdot\lr{q^i\bol{v}}=\nu C^{q^i},~~~~i=1,...,N.\label{Teq}
\end{equation}
where $C^{q^i}$ and $C^{p^i}$ are the diffusion (collision) operators associated with $q^i$ and $p^i$ (explicit expressions for the collision operators are given in section 3). Hence, each Clebsch potential can be thought of as a generalized distribution function that undergoes a relaxation process driven by viscosity.
Using \eqref{Teq}, we show that the Shannon entropy measure \cite{Shannon,Jaynes}  
\begin{equation}
{\rm H}\left[q^1,...,q^N\right]=-\sum_{i=1}^N \int_{\Omega}q^i\log q^i dV,\label{Hfunc}
\end{equation}
associated with the Clebsch potentials $q^i$, $i=1,...,N$, 
satisfies an H-theorem, i.e. the rate of change in ${\rm H}$ caused by the incompressible Navier-Stokes system \eqref{NS}, \eqref{BC} is non-negative:
\begin{equation}
\frac{d{\rm H}}{dt}\geq0.\label{dHdt0}
\end{equation}
Here, $dV$ is the volume element in $E^3$. 
The functional ${\rm H}$ can be interpreted as a measure of the topological complexity of the vortex structures associated with the  solution $\bol{v}$ at each instant $t$.
We also remark that, although both kinetic energy and the functional ${\rm H}$ satisfy an inequality fixing the sign of their growth rate, 
they are different from a thermodynamic standpoint. 
This difference is mathematically expressed by the fact that  
the functional ${\rm H}$ does not involve derivatives of the Clebsch potentials. 
A direct consequence of the inequality \eqref{dHdt0} is that equilibrium 
\begin{equation}
\bol{v}_{\infty}=\lim_{t\rightarrow\infty}\bol{v}\lr{\bol{x},t},\label{vinfty} 
\end{equation}
is vortex-free, 
\begin{equation}
\nabla\cp\bol{v}_{\infty}=\bol{0},
\end{equation}
provided that the limit \eqref{vinfty} is well defined.
Hence, $\bol{v}_{\infty}$ is independent of the initial data $\bol{v}_0$, and 
the potential $\phi_{\infty}$ such that $\bol{v}_{\infty}=\nabla\phi_{\infty}$  
can be obtained as a solution of the boundary value problem for Laplace's equation
\begin{equation}
\Delta\phi_{\infty}=0~~~~{\rm in}~~\Omega,~~~~\nabla\phi_{\infty}=\bol{0}~~~~{\rm on}~~\p\Omega,\label{Lap}
\end{equation}
arising from equations \eqref{NS2} and \eqref{BC}. 
However, the only smooth solutions of \eqref{Lap} are constants \cite{Kodaira49,Evans}, $\phi_{\infty}=c\in\mathbb{R}$, 
implying $\bol{v}_{\infty}=\bol{0}$.  
Notice that multivalued (angle) solutions for $\phi_{\infty}$ 
are not admissible because the value of $\nabla\phi_{\infty}$ 
is constrained both across and along the boundary $\p\Omega$. 

The fact that well-behaved solutions of system \eqref{NS}, \eqref{BC} decay 
as time advances can be made explicit by considering the 
budget equation for the kinetic energy \cite{Doering2},
\begin{equation}
\frac{1}{2}\frac{d}{dt}\int_{\Omega}\bol{v}^2 dV=-\nu\int_{\Omega}\bol{\omega}^2dV\leq 0.\label{dkdt}
\end{equation}
Here, $\bol{\omega}=\nabla\cp\bol{v}$ is the vorticity, 
and the boundary condition \eqref{BC} was used to eliminate surface integrals. 
The term on the right-hand side of equation \eqref{dkdt} is the enstrophy of the fluid.
If the limit $\bol{v}_{\infty}$ exists, each side of equation \eqref{dkdt} must vanish at equilibrium, 
implying $\nabla\cp\bol{v}_{\infty}=\bol{0}$ and therefore $\bol{v}_{\infty}=\bol{0}$. 
However, the time evolution of enstrophy is not trivial. 
Using the boundary condition \eqref{BC} and denoting with $\bol{n}$ the unit outward normal
to the bounding surface $\p\Omega$, 
it can be shown that
\begin{equation}
\frac{1}{2}\frac{d}{dt}\int_{\Omega}{\bol{\omega}}^2dV=\int_{\p\Omega}\nabla P\cdot\bol{\omega}\cp\bol{n}\,dS
-\int_{\Omega}\lr{\bol{v}\cdot\nabla\bol{v}}\cdot\nabla\cp\bol{\omega}\,dV-\nu\int_{\Omega}\abs{\nabla\cp\bol{\omega}}^2dV.
\end{equation} 
where $dS$ is the surface element on $\p\Omega$.
While the boundary term can be eliminated by suitable choice of boundary conditions (e.g. periodic
boundary conditions), the second (vortex stretching) term on the right-hand side does not have a definite sign.   
This term may lead to enstrophy growth. Currently, it is not known whether vortex stretching  
can produce finite time singularities in the solutions of the incompressible Navier-Stokes system  
\cite{Doering2,Ayala}. 
Vortex stretching does not arise in two dimensional flows, i.e. flows with velocity field $\bol{v}=\psi_y\nabla x-\psi_x\nabla y$
where $\psi\lr{x,y,t}$ is the stream function and the lower indexes denote partial derivatives.  
In such case, the rate of change in enstrophy is
\begin{equation}
\frac{1}{2}\frac{d}{dt}\int_{\Omega}\bol{\omega}^2dV=\nu\int_{\p\Omega}\omega\nabla\omega\cdot\bol{n}\,dS-\nu\int_{\Omega}\abs{\nabla\omega}^2dV,\label{dom2dt}
\end{equation}
with $\omega=-\Delta\psi$. 
This time the volume integral has a definite sign, implying a progressive decay of enstrophy whenever
the boundary term is zero. 
The decay rate of enstrophy in periodic two dimensional flows has been studied by Batchelor \cite{Batchelor}, who considered the high Reynolds number limit $\nu\rightarrow 0$ of equations \eqref{dkdt} and \eqref{dom2dt} 
and predicted by dimensional analysis \cite{Kolmogorov,Kraichnan} a decay law $\langle\bol{\omega}^2\rangle\sim t^{-2}$, where
$\langle\,\,\rangle$ denotes the spatial average. 
The decay of enstrophy for various Reynolds numbers has also been investigated numerically in periodic domains \cite{Chasnov,Dmitruk,Kida}, 
and a decay rate $\langle\bol{\omega}^2\rangle\sim t^{-0.8}$ has been observed in the limit of small viscosity. 

In the second part of the paper we introduce an alternative
Clebsch parametrization of the velocity field,
\begin{equation}
\bol{v}=\sum_{i=1}^R\nabla \mu^i\cp\nabla \lambda^i, \label{Clebsch_type2}
\end{equation}
with Clebsch parameters $\mu^i$, $\lambda^i$, $i=1,...,R$. 
This time, the Clebsch parametrization is complete if $\bol{v}$ is exact, i.e. it can be
expressed as $\bol{v}=\nabla\cp\bol{\xi}$ for some vector potential $\bol{\xi}$, and $R\geq 2$.
Using \eqref{Clebsch_type2}, we show that vortex stretching 
\begin{equation}
\int_{\Omega}\lr{\bol{v}\cdot\nabla\bol{v}}\cdot\nabla\times\bol{\omega}\,dV,\label{VS}
\end{equation}
identically vanishes whenever $R=1$, i.e. $\bol{v}=\nabla\mu\cp\nabla\lambda$ for some functions $\mu$ and $\lambda$. This result is a generalization of the two dimensional case where $\mu=\psi$, $\lambda=z$, and $\bol{v}=\nabla\psi\times\nabla z$. For such flows, a suitable choice of boundary conditions makes the rate of change in enstrophy non-positive.  





The present paper is organized as follows. 
Section 2 is a preliminary part that highlights   
the geometrical difficulty in sustaining nontrivial steady solutions $\bol{v}_{\infty}\neq\bol{0}$ of system \eqref{NS},  
and the tendency of time-dependent solutions to converge toward null equilibria $\bol{v}_{\infty}=\bol{0}$. In fact, by an integrability argument on the vorticity in \eqref{NS1}, 
one sees that all steady solutions of \eqref{NS} with $P+\bol{v}^2/2=c$, $c\in\mathbb{R}$, must be vortex-free.  
Furthermore, if the initial data $\bol{v}_0$ is given by a Beltrami field with constant proportionality coefficient \cite{Dombre86}, 
which is a steady solution of the ideal Euler equations expected to possess arbitrary topological complexity \cite{Arnold66,Enciso14,Enciso15}, it is  known that the resulting solution of \eqref{NS} is an exponentially decaying Beltrami field \cite{Drazin, YosRot}. 
In section 3 we write down system \eqref{NS} as a system of transport equations for the Clebsch parameters and identify the associated diffusion operators.
In section 4 we discuss the properties of the obtained diffusion operators, which depart from the standard Laplacian operator due to a term involving the Lie-bracket of the corresponding Clebsch pair.
In section 5 we show that the growth rate in the entropy measure ${\rm H}$ is always non-negative and deduce the vanishing of the corresponding equilibria, $\bol{v}_{\infty}=\bol{0}$. 
In section 6 we use the parametrization \eqref{Clebsch_type2} to show that, 
for flows of the type $\bol{v}=\nabla\mu\cp\nabla\lambda$, the vortex stretching term \eqref{VS} is zero, thus generalizing the classical result concerning two dimensional flows.  
Concluding remarks are given in section 7.

The results discussed in the present study do not represent a proof of existence of solutions to the incompressible Navier-Stokes equations.
They hold true provided that regular solutions of system \eqref{NS} exist in the limit $t\rightarrow\infty$.
For the Navier-Stokes existence and smoothness problem, we refer the reader to \cite{Ladyzhenskaya,Temam84,Doering}

Finally, in our analysis we use standard results of differential geometry.
We assume the reader to be familiar with the notions of integrability as described by the  
Frobenius \cite{Frankel} and Lie-Darboux theorems \cite{deLeon,Arnold89}.

\section{Geometrical Constraints on Field Topology in the Steady Incompressible Navier-Stokes Equations}

The purpose of the present section is to 
describe certain analytically tractable situations that result in the onset of null configurations $\bol{v}_{\infty}=\bol{0}$ in steady solutions of \eqref{NS}. 
First, we consider steady solutions of equation \eqref{NS} such that the Bernoulli head is constant,
\begin{equation}
P_{\infty}+\frac{\bol{v}^2_{\infty}}{2}=c,~~~~c\in\mathbb{R}.\label{BH}
\end{equation}
Here, we defined $P_{\infty}=\lim_{t\rightarrow\infty}P\lr{\bol{x},t}$. 
Although the constancy of the Bernoulli head is not justified in a general setting, this condition is what is expected for steady configurations where the nonlinear term in equation \eqref{NS1} balances dissipation.
Then, if a steady solution $\bol{v}_{\infty}$ of \eqref{NS} exists, it must satisfy the following equation:
\begin{equation}
\bol{v}_{\infty}\cp\bol{\omega}_{\infty}=\nu\nabla\cp\bol{\omega}_{\infty},\label{OmegaInfty}
\end{equation}
where $\bol{\omega}_{\infty}=\nabla\cp\bol{v}_{\infty}$. 
If $\bol{\omega}_{\infty}=\bol{0}$, the solution $\bol{v_{\infty}}$ is vortex-free. 
From the divergence-free condition \eqref{NS2} and the boundary condition \eqref{BC}, this implies 
that $\bol{v}_{\infty}=\nabla\phi_{\infty}$ must be a harmonic vector field with potential $\phi_{\infty}$ obeying
\begin{equation}
\Delta\phi_{\infty}=0~~~~{\rm in}~~\Omega,~~~~\nabla\phi_{\infty}=\bol{0}~~~~{\rm on}~~\p\Omega.\label{phiin}
\end{equation}
However, regular solutions of equation \eqref{phiin} are given by $\phi_{\infty}=c$, $c\in\mathbb{R}$. Therefore $\bol{v}_{\infty}=\bol{0}$. 
The fact that multivalued (angle) solutions for $\phi_{\infty}$ are not allowed can be seen as follows.  
Suppose that $\bol{v}_{\infty}=\bol{h}$ is a smooth harmonic vector field such that $\nabla\cp\bol{h}=\bol{0}$ and $\nabla\cdot\bol{h}=0$ in $\Omega$, 
and $\bol{h}=0$ on $\p\Omega$. Then, the Poincar\'e inequality implies that
\begin{equation}
\begin{split}
\Norm{\bol{h}}^2\leq C\sum_{i=1}^3\Norm{\nabla h_i}^2=&C\sum_{i=1}^3\lr{\int_{\p\Omega}h_i\nabla h_i\cdot\bol{n}\,dS-\int_{\Omega}h_i\Delta h_i\,dV}\\=&C\int_{\Omega}\bol{h}\cdot\left[\nabla\cp\lr{\nabla\cp\bol{h}}-\nabla\lr{\nabla\cdot\bol{h}}\right]\,dV=0.
\end{split}
\end{equation}
Here, $\Norm{\bol{h}}^2=\int_{\Omega}\bol{h}^2\,dV$ is the $L^2\lr{\Omega}$ norm of $\bol{h}$  
and $C\in\mathbb{R}$ a constant. 
Hence, $\bol{h}=\bol{0}$ in $\Omega$.

Now consider the case $\bol{\omega}_{\infty}\neq\bol{0}$. 
Multiplying each side of equation \eqref{OmegaInfty} by $\bol{\omega}_{\infty}$, it follows that the
helicity density of the vorticity $\bol{\omega}_{\infty}$ must vanish:
\begin{equation}
h_{\bol{\omega}_{\infty}}=\bol{\omega}_{\infty}\cdot\nabla\cp\bol{\omega}_{\infty}=0.\label{Frob}
\end{equation}
However, equation \eqref{Frob} is the Frobenius integrability condition \cite{Frankel} for the vector field $\bol{\omega}_{\infty}$:
assuming $\bol{\omega}_{\infty}$ to be smooth in the domain $\Omega$, for every point $\bol{x}\in\Omega$
there exist a neighborhood $U\subset\Omega$ centered at $\bol{x}$ and  
smooth functions $\lambda$ and $C$ defined in $U$ such that
\begin{equation}
\bol{\omega}_{\infty}=\lambda\nabla C~~~~{\rm in}~~U.\label{Frob2}
\end{equation}
Substituting equation \eqref{Frob2} in equation \eqref{OmegaInfty}, we arrive at the local condition
\begin{equation}
\lambda\bol{v}_{\infty}\cp\nabla C=\nu\nabla \lambda\cp\nabla C~~~~{\rm in}~~U.\label{Frob3}
\end{equation}
Equation \eqref{Frob3} implies that
\begin{equation}
\lambda\bol{v_{\infty}}=\nu\nabla\lambda+\mu\nabla C~~~~{\rm in}~~U,\label{Frob4}
\end{equation}
where $\mu$ is a function defined in $U$. Taking the curl of equation \eqref{Frob4}, multiplying it by $\lambda$, and using equation \eqref{Frob2},
we arrive at
\begin{equation}
\nabla\lambda\cp\lambda\bol{v}_{\infty}+\lambda^3\nabla C=\lambda\nabla\mu\cp\nabla C~~~~{\rm in}~~U.
\end{equation}
Substituting again \eqref{Frob4}, 
\begin{equation}
\lr{\mu\nabla\lambda-\lambda\nabla\mu}\cp\nabla C=-\lambda^3\nabla C~~~~{\rm in}~~U.
\end{equation}
Evidently, the left-hand side and right-hand side of this equation are orthogonal to each other. 
Hence, each side must vanish independently. In particular, the right-hand side vanishes 
if and only if $\bol{\omega}_{\infty}=\bol{0}$ in $U$. 
Since the argument applies to any neighborhood $U$ in $\Omega$, 
$\bol{v}_{\infty}=\nabla\phi_{\infty}$ in $\Omega$. 
In light of equation \eqref{phiin}, we arrive again at $\bol{v}_{\infty}=\bol{0}$.
This result shows that whenever the Bernoulli head $P+\bol{v}^2/2$ is constant, 
there are no nontrivial steady solutions of the incompressible Navier-Stokes equations \eqref{NS}.

Next, suppose that the initial data $\bol{v}_0$ for system \eqref{NS}
 is given by a Beltrami field with proportionality coefficient $\alpha\in\mathbb{R}$, $\alpha\neq 0$, 
\begin{equation}
\nabla\cp\bol{v}_{0}=\alpha\,\bol{v}_{0}.
\end{equation}
We also assume again that $P+\bol{v}^2/2=c$, $c\in\mathbb{R}$, for all $t\geq 0$. 
This setting is instructive because 
it is thought that Beltrami fields can produce vortex lines with desirable topological complexity (i.e. 
arbitrarily intricate links and knots) \cite{Arnold66}. 
Results in support of this conjecture can be found in \cite{Enciso15}. 
Hence, one may argue that knowing the outcome of \eqref{NS} for such an initial condition
is enough to constrain the outcome of a wider class of initial data $\bol{v}_0$ 
that can be `approximated' by a Beltrami field.
At $t=0$ we have:
\begin{equation}
\frac{\p\bol{v}}{\p t}\lr{\bol{x},0}=-\nu\alpha^2\bol{v}_0.
\end{equation}
Hence, the initial change in $\bol{v}$ is aligned with $\bol{v}_0$. In particular, at first order in $\epsilon>0$,
we may write:
\begin{equation}
\bol{v}\lr{\bol{x},\epsilon}=\lr{1-\epsilon\nu\alpha^2}\bol{v}_0+o\lr{\epsilon^2}=e^{-\epsilon\nu\alpha^2}\bol{v}_0+o\lr{\epsilon^2}.
\end{equation}
Repeating the argument $n=t/\epsilon$ times,  
\begin{equation}
\bol{v}\lr{\bol{x},n\epsilon}=e^{-n\epsilon\nu\alpha^2}\bol{v}_0+o\lr{\epsilon^2}.
\end{equation}
Taking the limit $\epsilon\rightarrow 0$ while keeping $t$ constant, we conclude that the solution of \eqref{NS} is
\begin{equation}
\bol{v}\lr{\bol{x},t}=e^{-\nu\alpha^2 t}\bol{v}_0
\end{equation}
Evidently, this solution tends to zero in the limit $t\rightarrow\infty$, implying a vortex-free null equilibrium $\bol{v}_{\infty}=\bol{0}$. 


\section{Clebsch Parametrization of the Incompressible Navier-Stokes Equations}

Consider the Clebsch parametrization \eqref{Clebsch0} of fluid velocity:
\begin{equation}
\bol{v}=\nabla\phi+\sum_{i=1}^N p^i\nabla q^i.\label{Clebsch}
\end{equation}
Here, $\phi\lr{\bol{x},t}$, $p^i\lr{\bol{x},t}$, $q^i\lr{\bol{x},t}$, $i=1,...,N$, are Clebsch potentials,
and $N\geq 2$ a natural number. 
Recall that in three spatial dimensions the parametrization \eqref{Clebsch} is complete, i.e. it is sufficient to represent
an arbitrary vector field. Furthermore, when $N\geq 3$ one can enforce boundary conditions on the Clebsch parameters. 
In the following, we shall refer to each pair $\lr{p^i,q^i}$ as a Clebsch pair, 
conjugated variables, or canonical variables. 

Substituting equation \eqref{Clebsch} into equation \eqref{NS1} and using standard vector identities, one obtains
\begin{equation}
\nabla\lr{\phi_t+\frac{\bol{v}^2}{2}+P+\sum_{i=1}^Np^iq^i_t}+\sum_{i=1}^N\left[\lr{p_t^i+\bol{v}\cdot\nabla p^i}\nabla q^i-\lr{q_t^i+\bol{v}\cdot\nabla q^i}\nabla p^i+\nu\nabla\cp\lr{\nabla p^i\cp\nabla q^i}\right]=\bol{0}.\label{NSC1}
\end{equation}  
Here, a lower index denotes partial differentiation, e.g. $\phi_t=\p\phi/\p t$. 
Next, recall the vector identity
\begin{equation}
\nabla\cp\lr{\nabla p^i\cp\nabla q^i}=\Delta q^i\nabla p^i-\Delta p^i\nabla q^i+\left[q^i,p^i\right],\label{VI1}
\end{equation}
where $\left[q^i,p^i\right]$ is a shorthand notation for the Lie bracket of vector fields
\begin{equation}
\left[q^i,p^i\right]=\lr{\nabla q^i\cdot\nabla}\nabla p^i-\lr{\nabla p^i\cdot\nabla}\nabla q^i.
\end{equation}
Substituting again equation \eqref{VI1} into equation \eqref{NSC1}, we find
\begin{equation}
\begin{split}
\nabla&\lr{\phi_t+\frac{\bol{v}^2}{2}+P+\sum_{i=1}^Np^iq^i_t}=\\
&\sum_{i=1}^N\left[\lr{q_t^i+\bol{v}\cdot\nabla q^i-\nu\Delta q^i}\nabla p^i-\lr{p_t^i+\bol{v}\cdot\nabla p^i-\nu\Delta p^i}\nabla q^i+\nu\left[p^i,q^i\right]\right].\label{NSC2}
\end{split}
\end{equation}
At this point, we distinguish three cases. 

\underline{Case 1}: the $i$th component $\bol{\omega}^i=\nabla p^i\cp\nabla q^i$ of the vorticity $\bol{\omega}=\nabla\cp\bol{v}=\sum_{i=1}^N\bol{\omega}^i$ is zero, $\bol{\omega}^i=\bol{0}$. 
Then, $p^i$ is a function of $q^i$, and the term $p^i\nabla q^i$
in the parametrization \eqref{Clebsch} can be absorbed in the Clebsch potential $\phi$, effectively reducing by one the number
of Clebsch pairs $N$. 

\underline{Case 2}: the $i$th component of the vorticity is different from zero, $\bol{\omega}^i\neq\bol{0}$, but its helicity density
\begin{equation}
h_{\bol{\omega}^i}=\bol{\omega}^i\cdot\nabla\cp\bol{\omega}^i=\nabla\cp\lr{\nabla p^i\cp\nabla q^i}\cdot\nabla p^i\cp\nabla q^i=-\left[p^i,q^i\right]\cdot\nabla p^i\cp\nabla q^i,\label{helicity}
\end{equation}
vanishes, $h_{\bol{\omega}^i}=0$. Then, the term $\left[p^i,q^i\right]$ on the right-hand side
of equation \eqref{NSC2} can be decomposed by looking for a solution $\eta^i$ of the following first order
partial differential equation:
\begin{equation}
\nabla\eta^i\cdot\nabla p^i\cp\nabla q^i=\abs{\nabla p^i\cp\nabla q^i}.\label{Jac2}
\end{equation} 
Defining the curvilinear coordinate system $\lr{p^i,q^i,\eta^i}$, we have
\begin{equation}
\begin{split}
\left[p^i,q^i\right]=&\frac{\left[p^i,q^i\right]\cdot\nabla p^i\cp\nabla q^i}{\nabla\eta^i\cdot\nabla p^i\cp\nabla q^i}\nabla\eta^i+\frac{\left[p^i,q^i\right]\cdot\nabla q^i\cp\nabla \eta^i}{\nabla\eta^i\cdot\nabla p^i\cp\nabla q^i}\nabla p^i
+\frac{\left[p^i,q^i\right]\cdot\nabla \eta^i\cp\nabla p^i}{\nabla\eta^i\cdot\nabla p^i\cp\nabla q^i}\nabla q^i\\
=&\frac{\left[p^i,q^i\right]\cdot\nabla q^i\cp\nabla \eta^i}{\abs{\nabla p^i\cp\nabla q^i}}\nabla p^i
+\frac{\left[p^i,q^i\right]\cdot\nabla \eta^i\cp\nabla p^i}{\abs{\nabla p^i\cp\nabla q^i}}\nabla q^i.\label{LB2}
\end{split}
\end{equation} 
In deriving this equation we used the hypothesis $h_{\bol{\omega}^i}=0$, equation \eqref{Jac2}, and the fact that 
in a curvilinear coordinate system $\lr{x^1,x^2,x^3}$ a vector field $\bol{w}$ can be decomposed
along the cotangent basis $\lr{\nabla x^1,\nabla x^2,\nabla x^3}$ according to $\bol{w}=\lr{\bol{w}\cdot\p_i}\nabla x^i$,
with $\p_i=\epsilon_{ijk}\nabla x^j\cp\nabla x^k/\nabla x^i\cdot\nabla x^j\cp\nabla x^k$ the $i$th tangent vector.

\underline{Case 3}: the $i$th component of the vorticity and its helicity density are both different from zero, $\bol{\omega}^i\neq\bol{0}$
and $h_{\bol{\omega}^i}\neq 0$. Now the term $\left[p^i,q^i\right]$ on the right-hand side
of equation \eqref{NSC2} can be decomposed by finding a solution $\xi^i$ of the following first order
partial differential equation:
\begin{equation}
\nabla\xi^i\cdot\nabla p^i\cp\nabla q^i=-h_{\bol{\omega^i}},\label{Jac3}
\end{equation} 
This definition for $\xi^i$ was introduced in reference \cite{Scholle16} for the case $N=1$. 
Then, defining the curvilinear coordinate system $\lr{p^i,q^i,\xi^i}$, we have
\begin{equation}
\begin{split}
\left[p^i,q^i\right]=&\frac{\left[p^i,q^i\right]\cdot\nabla p^i\cp\nabla q^i}{\nabla\xi^i\cdot\nabla p^i\cp\nabla q^i}\nabla\xi^i+\frac{\left[p^i,q^i\right]\cdot\nabla q^i\cp\nabla \xi^i}{\nabla\xi^i\cdot\nabla p^i\cp\nabla q^i}\nabla p^i
+\frac{\left[p^i,q^i\right]\cdot\nabla \xi^i\cp\nabla p^i}{\nabla\xi^i\cdot\nabla p^i\cp\nabla q^i}\nabla q^i\\
=&\nabla\xi^i+\frac{\left[p^i,q^i\right]\cdot\nabla q^i\cp\nabla \xi^i}{\left[p^i,q^i\right]\cdot\nabla p^i\cp\nabla q^i}\nabla p^i
+\frac{\left[p^i,q^i\right]\cdot\nabla \xi^i\cp\nabla p^i}{\left[p^i,q^i\right]\cdot\nabla p^i\cp\nabla q^i}\nabla q^i.\label{LB3}
\end{split}
\end{equation} 
In the last passage, we used equation \eqref{helicity}. 

For simplicity, we assume that each component $\bol{\omega}^i$, $i=1,...,N$,
belongs to one of the three cases listed above in the whole of $\Omega$. 
However, in general the components $\bol{\omega}^i$ may behave differently
in different regions of $\Omega$. 
In such case, each region must be treated independently. 

Let $N$ be the number of nontrivial Clebsch pairs such that $\bol{\omega}^i\neq\bol{0}$, $i=1,...,N$. 
Let $M=N-R$ be the number of Clebsch pairs whose vorticity $\bol{\omega}^i$, $i=1,...,M$, has vanishing helicity density $h_{\bol{\omega}^i}=0$ and $R=N-M$ the number of Clebsch pairs whose vorticity $\bol{\omega}^i$, $i=M+1,...,N$, has non-vanishing 
helicity density $h_{\bol{\omega}^i}\neq0$. 
Substituting the decompositions \eqref{LB2} and \eqref{LB3} into equation \eqref{NSC2}, we arrive at the following form for equation \eqref{NS1}:
\begin{equation}
\begin{split}
\nabla\left[\phi_t+\frac{\bol{v}^2}{2}+P+\sum_{i=1}^Np^iq^i_t-\nu\sum_{i=M+1}^N\xi^i\right]=&
\sum_{i=1}^M\left[q_t^i+\bol{v}\cdot\nabla q^i-\nu\lr{\Delta q^i-\frac{\left[p^i,q^i\right]\cdot\nabla q^i\cp\nabla\eta^i}{\abs{\nabla p^i\cp\nabla q^i}}}\right]\nabla p^i
\\&-\sum_{i=1}^M\left[p_t^i+\bol{v}\cdot\nabla p^i-\nu\lr{\Delta p^i+\frac{\left[p^i,q^i\right]\cdot\nabla\eta^i\cp\nabla p^i}{\abs{\nabla p^i\cp\nabla q^i}}}\right]\nabla q^i\\&+
\sum_{i=M+1}^N\left[q_t^i+\bol{v}\cdot\nabla q^i-\nu\lr{\Delta q^i-\frac{\left[p^i,q^i\right]\cdot\nabla q^i\cp\nabla\xi^i}{\left[p^i,q^i\right]\cdot\nabla p^i\cp\nabla q^i}}\right]\nabla p^i
\\&-\sum_{i=M+1}^N\left[p_t^i+\bol{v}\cdot\nabla p^i-\nu\lr{\Delta p^i+\frac{\left[p^i,q^i\right]\cdot\nabla\xi^i\cp\nabla p^i}{\left[p^i,q^i\right]\cdot\nabla p^i\cp\nabla q^i}}\right]\nabla q^i.\label{NSC3}
\end{split}
\end{equation}
Evidently, the original Navier-Stokes system \eqref{NS} can be satisfied by finding Clebsch potentials $\phi$, $p^i$, $q^i$, $i=1,...,N$, $\eta^j$, $j=1,...,M$, $\xi^k$, $k=M+1,...,N$, and a pressure $P$ satisfying the following system of $3N+2$ equations
\begin{subequations}\label{NSC4}
\begin{align}
P=&-\phi_t-\frac{\bol{v}^2}{2}-\sum_{i=1}^N p^iq_t^i+\nu\sum_{i=M+1}^N\xi^i,\label{Pt}\\
q_t^i=&-\nabla\cdot\lr{\bol{v}q^i}+\nu\lr{\Delta q^i-\frac{\left[p^i,q^i\right]\cdot\nabla q^i\cp\nabla\eta^i}{\abs{\nabla p^i\cp\nabla q^i}}},~~~~i=1,...,M,\label{qt1}\\
p_t^i=&-\nabla\cdot\lr{\bol{v}p^i}+\nu\lr{\Delta p^i+\frac{\left[p^i,q^i\right]\cdot\nabla\eta^i\cp\nabla p^i}{\abs{\nabla p^i\cp\nabla q^i}}},~~~~i=1,...,M,\label{pt1}\\
q_t^i=&-\nabla\cdot\lr{\bol{v}q^i}+\nu\lr{\Delta q^i-\frac{\left[p^i,q^i\right]\cdot\nabla q^i\cp\nabla\xi^i}{\left[p^i,q^i\right]\cdot\nabla p^i\cp\nabla q^i}},~~~~i=M+1,...,N,\label{qt2}\\
p_t^i=&-\nabla\cdot\lr{\bol{v}p^i}+\nu\lr{\Delta p^i+\frac{\left[p^i,q^i\right]\cdot\nabla\xi^i\cp\nabla p^i}{\left[p^i,q^i\right]\cdot\nabla p^i\cp\nabla q^i}},~~~~i=M+1,...,N,\label{pt2}\\
\nabla \eta^i\cdot\nabla p^i\cp\nabla q^i=&\abs{\nabla p^i\cp\nabla q^i},~~~~i=1,...,M,\\
\nabla \xi^i\cdot\nabla p^i\cp\nabla q^i=&\left[p^i,q^i\right]\cdot\nabla p^i\cp\nabla q^i,~~~~i=M+1,...,N,\\
\Delta\phi=&-\sum_{i=1}^N\nabla\cdot\lr{p^i\nabla q^i}\label{phit}.
\end{align}
\end{subequations}
Here, the last equation is the divergence-free condition \eqref{NS2}.   
On the other hand, recall that the Clebsch parametrization \eqref{Clebsch} is complete. 
Hence, any initial condition $\bol{v}_0=\bol{v}\lr{\bol{x},0}$ can be expressed by appropriate choice of initial conditions on 
the Clebsch potentials, and the corresponding solution of system \eqref{NSC4} at time $t$ provides a solution $\bol{v}\lr{\bol{x},t}$ of the incompressible Navier-Stokes system \eqref{NS} with initial condition $\bol{v}_0$. 

Observe that when $\nu=0$ system \eqref{NSC4} gives the canonical Hamiltonian form of the incompressible ideal Euler equations of fluid dynamics (see \cite{YosEpi2D}).
For this reason, we shall refer to system \eqref{NSC4} as the canonical form of the incompressible Navier-Stokes system \eqref{NS} (notice that, however, the Navier-Stokes equations are not a Hamiltonian system due to the viscous term).

We further remark that in this construction it is assumed that the decompositions of cases 1, 2, and 3 hold during
time evolution throughout the region $\Omega$, 
i.e. a Clebsch pair without vorticity cannot transition to a state with finite vorticity and viversa. Similarly,  
a Clebsch pair with helicity density $h_{\bol{\omega}^i}=0$ cannot transition to a state with helicity density $h_{\bol{\omega}^i}\neq0$
and viceversa. To understand the meaning of this hypothesis, it is useful to explain the geometrical constraint posed by the condition $h_{\bol{\omega}^i}=0$. This requirement is nothing but the Frobenius integrability condition \cite{Frankel} for the vector field $\bol{\omega}^i$, i.e. the mathematical condition that must be satisfied by the vorticity $\bol{\omega}^i$ for the corresponding vortex
lines to define the normal direction of a locally defined surface in $\Omega$. More precisely, if $\bol{h}_{\bol{\omega}^i}=0$ in $\Omega$, for every point $\bol{x}\in\Omega$ there exists a neighborhood $U\subset\Omega$ of $\bol{x}$ and smooth functions
$\lambda^i$ and $C^i$ defined in $U$ such that
\begin{equation}
\bol{\omega}^i=\lambda^i\nabla C^i~~~~{\rm in}~~U,
\end{equation}
so that $\bol{\omega}^i$ is aligned with the normal to the local surface $C={\rm constant}$. 
The situation in which $h_{\bol{\omega}^i}=0$ thus represents a generalization of two dimensional incompressible fluid flows.
Indeed, denoting with $\psi\lr{x,y,t}$ the stream function, the velocity field of a two dimensional incompressible flow in the $\lr{x,y}$ plane can be expressed as
\begin{equation}
\bol{v}=\psi_y\nabla x-\psi_x\nabla y.\label{v2D}
\end{equation} 
Then, the vorticity of the system is
\begin{equation}
\bol{\omega}=-\Delta\psi\nabla z,\label{omega2D}
\end{equation}  
so that the helicity density of the vorticity is $h_{\bol{\omega}}=\bol{\omega}\cdot\nabla\cp\bol{\omega}=0$. 
This implies $\bol{\omega}=\lambda\nabla C$ with $\lambda=-\Delta\psi$ and $C=z$. 
Therefore, the hypothesis that the integrability properties of each vorticity component $\bol{\omega}^i$ are preserved during time evolution
can be understood as the requirement that transitions from two to three dimensional behavior and viceversa are not admitted. 
In a similar way, the condition that each vorticity component $\bol{\omega}^i$ cannot transition from a non-zero value to zero and viceversa
signifies that the topological structure of the vorticity $\bol{\omega}$ is conserved in time. 
If transitions of the type described above are instead allowed, 
one must adjust the numbers $N$, $M$, $R$ and the corresponding
variables $\eta^i$, $i=1,...,M$, and $\xi^i$, $i=M+1,...,N$ during time evolution.

\section{Properties of the Diffusion Operator for the Clebsch Potentials}
The evolution equations \eqref{qt1}, \eqref{pt1}, \eqref{qt2} and \eqref{pt2} satisfied by the Clebsch potentials $p^i$ and $q^i$, $i=1,...,N$, can be regarded as
transport equations where the convective part is driven by the velocity field $\bol{v}$ while diffusion 
is scaled by the diffusion parameter $\nu$ with diffusion operators
\begin{subequations}\label{DO}
\begin{align}
C^{q^i}=&\Delta q^i-\frac{\left[p^i,q^i\right]\cdot\nabla q^i\cp\nabla\eta^i}{\abs{\nabla p^i\cp\nabla q^i}},~~~~
C^{p^i}=\Delta p^i+\frac{\left[p^i,q^i\right]\cdot\nabla\eta^i\cp\nabla p^i}{\abs{\nabla p^i\cp\nabla q^i}},~~~~i=1,...,M.\label{DO1}\\
C^{q^i}=&\Delta q^i-\frac{\left[p^i,q^i\right]\cdot\nabla q^i\cp\nabla\xi^i}{\left[p^i,q^i\right]\cdot\nabla p^i\cp\nabla q^i},~~~~
C^{p^i}=\Delta p^i+\frac{\left[p^i,q^i\right]\cdot\nabla\xi^i\cp\nabla p^i}{\left[p^i,q^i\right]\cdot\nabla p^i\cp\nabla q^i},~~~~i=M+1,...,N.\label{DO2}
\end{align}
\end{subequations}
These diffusion operators depart from the standard Laplacian diffusion operator $\Delta$ due to the terms involving 
the Lie bracket $\left[p^i,q^i\right]$, highlighting the non-triviality of the dissipation process associated with the incompressible Navier-Stokes system \eqref{NS}. 
In particular, notice that the diffusion operator of each Clebsch potential $q^i$ is not determined by $q^i$ alone, but it is coupled with the behavior of the conjugated variable $p^i$ and viceversa.

It is instructive to derive the expressions of the diffusion operators \eqref{DO} for two dimensional incompressible flows.  
For a two dimensional vector field, the parametrization \eqref{Clebsch} is complete for $N\geq 1$. We set $N=1$, $p=p^1$, $q=q^1$, and $\bol{v}=\nabla\phi+p\nabla q$.
Notice that the Clebsch potentials are independent of $z$, i.e. $\phi=\phi\lr{x,y,t}$, $p=p\lr{x,y,t}$, and $q=q\lr{x,y,t}$.
Furthermore, the helicity density of the vorticity is zero (recall equation \eqref{omega2D}). Hence, this system falls in case 2 of the previous section, 
and the diffusion operators of interest are those in equation \eqref{DO1} with $\eta=\eta^1$ given by a solution of
\begin{equation}
-\Delta\psi\nabla\eta\cdot\nabla z=\abs{\Delta\psi}.\label{eta2D}
\end{equation} 
Here, we used equations \eqref{Jac2} and \eqref{omega2D}.
Suppose that $\Delta\psi<0$ (vorticity is positive). Then a solution of \eqref{eta2D} can be obtained by setting $\eta=z$. 
It follows that the diffusion operators \eqref{DO1} for the Clebsch potentials $p$ and $q$ can be written as
\begin{equation}
C^q=\nabla q\cdot\nabla\log\abs{\nabla p\cp\nabla q},~~~~C^p=\nabla p\cdot\log\abs{\nabla p\cp\nabla q}.
\end{equation}
Hence, in two dimensions, the canonical incompressible Navier-Stokes system \eqref{NSC4} becomes:
\begin{subequations}
\begin{align}
P=&-\phi_t-\frac{\bol{v}^2}{2}-pq_t,\\
q_t=&-\nabla\cdot\lr{\bol{v}q}+\nu\nabla q\cdot\nabla\log\abs{\nabla p\cp\nabla q},\\
p_t=&-\nabla\cdot\lr{\bol{v}p}+\nu\nabla p\cdot\nabla\log\abs{\nabla p\cp\nabla q},\\
\Delta\phi=&-\nabla\cdot\lr{p\nabla q}.
\end{align}
\end{subequations}

The remaining part of this section is dedicated to the derivation of an alternative form for the diffusion operators \eqref{DO}.  
This form will be used in the proof of the H-theorem given in the next section. 
Let $\alpha$ denote one of the Clebsch potentials $p^i$, $q^i$, $i=1,...,N$. 
From \eqref{DO}, we see that the corresponding diffusion operator can be written as
\begin{equation}
C^{\alpha}=\Delta\alpha+\nabla\alpha\cdot\bol{w}^\alpha,\label{DOalpha} 
\end{equation}
with $\bol{w}^{\alpha}$ a vector field. 
We wish to show that there exist a vector potential $\bol{A}^{\alpha}$ 
with the property that 
\begin{equation}
\bol{w}^{\alpha}=\nabla\cp\bol{A}^{\alpha}
.\label{Rep1}
\end{equation}
Let us consider the case $C^{q^i}$ in \eqref{DO2}, the other cases being analogous. 
The vector $\left[p^i,q^i\right]$ appearing in the definition of the diffusion operator 
can be decomposed on the cotangent basis $\lr{\nabla p^i,\nabla q^i,\nabla \xi^i}$ as \begin{equation}
\left[p^i,q^i\right]=f_{p^i}\nabla p^i+f_{q^i}\nabla q^i+f_{\xi^i}\nabla\xi^i, 
\end{equation}
for appropriate coefficients $f_{p^i}$, $f_{q^i}$, and $f_{\xi^i}$. 
It follows that
\begin{equation}
C^{q^i}=\Delta q^i+\frac{f_{p^i}}{h_{\bol{\omega}^i}}\nabla q^i\cdot\nabla \xi^i\cp\nabla p^i=\Delta q^i+\nabla q^i\cdot\nabla\lr{\int\frac{f_{p^i}}{h_{\bol{\omega}}^i}d\xi^i}\cp\nabla p^i.
\end{equation}
Hence, defining the quantity 
\begin{equation}
\Theta^i=\int \frac{f_{p^i}}{h_{\bol{\omega^i}}}d\xi^i=-\int\frac{\left[p^i,q^i\right]\cdot\nabla q^i\cp\nabla \xi^i}{\lr{\nabla\xi^i\cdot\nabla p^i\cp\nabla q^i}^2}d\xi^i,
\end{equation}
we may set
\begin{equation}
\bol{A}^{q^i}=\Theta^i\nabla p^i.
\end{equation}
The diffusion operator \eqref{DOalpha} has thus expression
\begin{equation}
C^{\alpha}=\Delta\alpha+\nabla\alpha\cdot\nabla\cp\bol{A}^{\alpha}=\nabla\cdot\lr{\nabla\alpha+\bol{A}^{\alpha}\cp\nabla\alpha}.
\end{equation}
Furthermore, the corresponding evolution equation for the Clebsch potential $\alpha$ is
\begin{equation}
\alpha_t=-\nabla\cdot\lr{\bol{V}^\alpha\alpha},~~~~\bol{V}^{\alpha}=\bol{v}-\nu\lr{\nabla\log\alpha+\bol{A}^{\alpha}\cp\nabla\log\alpha},\label{alphat}
\end{equation}
with $\bol{V}^\alpha$ the effective fluid velocity advecting the potential $\alpha$.  

It is useful to workout the expression of $\nabla\cp\bol{A}^{\alpha}$ for one of the diffusion operators, say $C^{q}$,
for the two dimensional case discussed above. From equation \eqref{DO1}, and recalling that 
$N=1$ with $\eta=z$, we have
\begin{equation}
\begin{split}
C^q=&\Delta q-\frac{\nabla q\cdot\nabla z\cp\left[p,q\right]}{\abs{\nabla p\cp\nabla q}}\\
=&\Delta q-\nabla q\cdot\left[\nabla\cp\lr{\frac{z\left[p,q\right]}{\abs{\nabla p\cp\nabla q}}}-z\nabla\cp\lr{\frac{\left[p,q\right]}{\abs{\nabla p\cp\nabla q}}}\right]\\
=&\Delta q-\nabla q\cdot\left[\nabla\cp\lr{\frac{z\left[p,q\right]}{\abs{\nabla p\cp\nabla q}}}\right]+z\nabla\cdot\lr{\frac{\left[p,q\right]\cp\nabla q}{\abs{\nabla p\cp\nabla q}}}.
\end{split}
\end{equation}
Since in two dimensions $p=p\lr{x,y,t}$, $q=q\lr{x,y,t}$, and $\bol{\omega}=\nabla p\cp\nabla q=-\Delta\psi\nabla z$ with $\psi\lr{x,y,t}$ the stream function, 
one can verify that
\begin{equation}
\nabla\cdot\lr{\frac{\left[p,q\right]\cp\nabla q}{\abs{\nabla p\cp\nabla q}}}=0.
\end{equation}
It follows that
\begin{equation}
C^q=\Delta q-\nabla q\cdot\left[\nabla\cp\lr{\frac{z\left[p,q\right]}{\abs{\nabla p\cp\nabla q}}}\right],
\end{equation}
and 
\begin{equation}
\nabla\cp\bol{A}^{q}=-\nabla\cp\lr{\frac{z\left[p,q\right]}{\abs{\nabla p\cp\nabla q}}}.
\end{equation}

\section{Entropy Measures and Relaxation}

An H-theorem is a statement regarding the non-decreasing nature of a functional ${\rm H}$ measuring
a certain type of order in a physical system subject to thermodynamically irreversible processes. 
The functional ${\rm H}$ can be though of as an information or entropy measure.
In the present context, recalling the analogy with transport processes, we are interested in defining a Shannon-type entropy   measure ${\rm H}$ reflecting the topological complexity
of the velocity field $\bol{v}$, and show that
\begin{equation}
\frac{d{\rm H}}{dt}\geq 0~~~~{\forall} t\geq0.\label{dHdt}
\end{equation}
This inequality then implies a progressive degradation in the topological complexity
of the fluid flow, and, if steady regular solutions of the incompressible Navier-Stokes system exist
in the limit $t\rightarrow\infty$, a null equilibrium $\bol{v}_{\infty}=\lim_{t\rightarrow\infty}\bol{v}=\bol{0}$.    

Notice that other physical quantities may obey analogous inequalities. 
In particular, the energy $E=\frac{1}{2}\int_{\Omega}{\bol{v}^2}dV$ decays over time
in the presence of a finite viscosity $\nu$. However, the roles played by entropy and energy
are different from a thermodynamic 
point of view. 

As explained in the previous section, the evolution equation for the Clebsch potentials \eqref{alphat}
can be interpreted as a transport equation where each Clebsch potential behaves as a generalized distribution function. 
Furthermore, due to the regularity hypothesis and the non-blowing up condition on the Clebsch parameters, each $q^i$ attains 
its minimum $q^i_m=q^i\lr{\bol{x}^i_m,t_m^i}$ at some point $\lr{\bol{x}^i_m,t_m^i}$, $i=1,...,N$. Then,  
the transformation $q^i\rightarrow q^i+\abs{q^i_m}+\epsilon^i$, with $\epsilon^i>0$, ensures that $q^i>0$ while leaving 
the velocity field \eqref{Clebsch} unchanged. This transformation is admissible because each $q^i$ appears only through temporal and spatial derivatives in system \eqref{NSC4}.
Therefore, if $q^i$ is a solution, so is $q^i+\abs{q^i_m}+\epsilon^i$.    
We therefore postulate that the functional ${\rm H}$ associated with the incompressible Navier-Stokes system \eqref{NS} is a Shannon-type information measure for  
positive Clebsch potentials $q^i$, $i=1,...,N$, defined as:
\begin{equation}
{\rm H}\left[q^1,...,q^N\right]=-\sum_{i=1}^N\int_{\Omega}q^i\log q^idV.
\end{equation} 
Denoting with 
\begin{equation}
\bol{V}^{q^i}=\bol{v}-\nu\lr{\nabla\log q^i+\bol{A}^{q^i}\cp\nabla\log q^i},\label{Vqi}
\end{equation} 
the effective fluid velocity associated with the Clebsch parameter $q^i$ 
as derived in equation \eqref{alphat}, we have
\begin{equation}
\begin{split}
\frac{d{\rm H}}{dt}=&-\sum_{i=1}^N\int_{\Omega}q^i_t\lr{1+\log q^i}dV\\
=&\sum_{i=1}^N\int_{\Omega}\nabla\cdot\lr{\bol{V}^{q^i}q^i}\lr{1+\log q^i}dV\\
=&\sum_{i=1}^N\left[\int_{\p\Omega}q^i\lr{1+\log q^i}\bol{V}^{q^i}\cdot\bol{n}\,dS-\int_{\Omega}q^i\bol{V}^{q^i}\cdot\nabla\log q^i dV\right]\\
=&\sum_{i=1}^N\left[\int_{\p\Omega}q^i\lr{1+\log q^i}\bol{V}^{q^i}\cdot\bol{n}\,dS-\int_{\Omega}\nabla\cdot\lr{\bol{v}q^i}dV+\nu\int_{\Omega}\lr{\nabla q^i+\bol{A}^{q^i}\cp\nabla q^i}\cdot\nabla\log q^i dV\right]\\
=&\sum_{i=1}^N\left\{\int_{\p\Omega}q^i\left[\lr{1+\log q^i}\bol{V}^{q^i}-\bol{v}\right]\cdot\bol{n}\,dS+\nu\int_{\Omega}q^i\abs{\nabla\log q^i}^2 dV\right\}\\
=&\sum_{i=1}^N\left\{-\nu\int_{\p\Omega}q^i\left[\lr{1+\log q^i}\lr{\nabla\log q^i+\bol{A}^{q^i}\cp\nabla\log q^i}\right]\cdot\bol{n}\,dS+\nu\int_{\Omega}q^i\abs{\nabla\log q^i}^2 dV\right\}\\
=&\sum_{i=1}^N\left\{-\nu\int_{\p\Omega}\left[\nabla\lr{q^i\log q^i}+\bol{A}^{q^i}\cp\nabla\lr{q^i\log q^i}\right]\cdot\bol{n}\,dS+\nu\int_{\Omega}q^i\abs{\nabla\log q^i}^2dV\right\}.\label{dHdt2}
\end{split}
\end{equation}
In these passages, we used the divergence theorem, equations \eqref{alphat} and \eqref{Vqi}, and the boundary condition 
$\bol{v}=\bol{0}$ on $\p\Omega$ (equation \eqref{BC}). In this notation, $\bol{n}$ represents the unit outward normal to the
bounding surface $\p\Omega$. 
Taking $N\geq 3$, there is enough freedom to enforce boundary conditions on the Clebsch potentials. 
In particular, we demand that the Clebsch potentials $\phi$, $q^i$, and $p^i$, $i=1,...,N$, satisfy the single boundary condition
\begin{equation}
\sum_{i=1}^N\int_{\p\Omega}\left[\nabla\lr{q^i\log q^i}+\bol{A}^{q^i}\cp\nabla\lr{q^i\log q^i}\right]\cdot\bol{n}\,dS=0.\label{BCClebsch}
\end{equation}
Equation \eqref{BCClebsch} can be written as a Neumann boundary condition for the Clebsch potential $\phi$. 
Indeed, from \eqref{Clebsch} and the boundary condition \eqref{BC} we may write 
\begin{equation}
p^1\nabla q^1=-\nabla\phi-\sum_{i=2}^Np^i\nabla q^i~~~~{\rm on}~~\p\Omega.
\end{equation} 
Then, equation \eqref{BCClebsch} can be satisfied by demanding that
\begin{equation}
\nabla\phi\cdot\bol{n}=\left[p^1\frac{\sum_{i=2}^N\nabla\lr{q^i\log q^i}+\sum_{i=1}^N\bol{A}^{q^i}\times\nabla\lr{q^i\log q^i}}{1+\log q^1}-\sum_{i=2}^Np^i\nabla q^i\right]\cdot\bol{n}~~~~{\rm on}~~\p\Omega.\label{BCphi}
\end{equation}
Hence, the Neumann boundary condition \eqref{BCphi} is the boundary condition to be enforced when solving
equation \eqref{phit} for the Clebsch potential $\phi$.


Using the boundary condition \eqref{BCphi}, equation \eqref{dHdt2} becomes
\begin{equation}
\frac{d{\rm H}}{dt}=\nu\sum_{i=1}^N\int_{\Omega}q^i\abs{\nabla\log q^i}^2dV\geq 0.\label{dHdt3}
\end{equation}
Here, we used the fact that, by construction, $q^i>0$, $i=1,...,N$. 
Equation \eqref{dHdt3} is the desired H-theorem. 

Let us make some considerations on the implications of the inequality \eqref{dHdt3}. 
Suppose that a regular solution $\lr{\phi,p^1,...,p^N,q^1,...,q^N,P}$ of system \eqref{NSC4} exists in the limit
$t\rightarrow\infty$. Then, the ${\rm H}$ function must approach a maximum, implying 
\begin{equation}
\lim_{t\rightarrow\infty}\frac{d{\rm H}}{dt}=0.\label{dHdt4}
\end{equation}
Since each $q^i$ is positive, in light of the expression of $d{\rm H}/dt$ in \eqref{dHdt3},  
equation \eqref{dHdt4} can be satisfied if and only if
\begin{equation}
\lim_{t\rightarrow\infty}\abs{\nabla\log q^i}=0,~~~~i=1,...,N.
\end{equation}
We therefore deduce that
\begin{equation}
\bol{v}_{\infty}=\lim_{t\rightarrow\infty}\bol{v}=\nabla \phi_{\infty},
\end{equation}
with $\phi_{\infty}=\lim_{t\rightarrow\infty}\phi$. 
On the other hand, such $\bol{v}_{\infty}$ is a steady solution of the 
incompressible Navier-Stokes system \eqref{NS} provided that
\begin{equation}
\Delta\phi_{\infty}=0~~~~{\rm in}~~\Omega,~~~~\nabla\phi_{\infty}=\bol{0}~~~~{\rm on}~~\p\Omega.\label{phiinfty} 
\end{equation}
As already seen, regular solutions of \eqref{phiinfty} are given by $\phi_{\infty}=c$ with $c\in\mathbb{R}$. 
We conclude that
\begin{equation}
\bol{v}_{\infty}=\bol{0}.
\end{equation}

It should be noted that it is possible to define alternative 
entropy measures. For example, the functionals
\begin{equation}
\mc{H}^q=-\frac{1}{2}\sum_{i=1}^N\int_{\Omega}\lr{q^i}^2dV,
~~~~\mc{H}^p=-\frac{1}{2}\sum_{i=1}^N\int_{\Omega}\lr{p^i}^2dV,
\end{equation}
can be shown to satisfy the budget equations
\begin{equation}
\frac{d\mc{H}^q}{dt}=\nu\sum_{i=1}^N\int_{\Omega}\abs{\nabla q^i}^2dV\geq 0,
~~~~\frac{d\mc{H}^p}{dt}=\nu\sum_{i=1}^N\int_{\Omega}\abs{\nabla p^i}^2dV\geq 0,
\end{equation}
provided that the Clebsch parameters are chosen so that
they satisfy the boundary conditions
\begin{equation}
\sum_{i=1}^N\int_{\Omega}{\lr{\nabla \lr{q^i}^2+\bol{A}^{q^i}\cp\nabla\lr{q^i}^2}\cdot\bol{n}\,dS}=0,~~~~
\sum_{i=1}^N\int_{\Omega}{\lr{\nabla \lr{p^i}^2+\bol{A}^{p^i}\cp\nabla\lr{p^i}^2}\cdot\bol{n}\,dS}=0
~~~~{\rm on}~~\p\Omega,\label{BCH}
\end{equation}
which, again, can be written as Neumann boundary conditions for the Clebsch potential $\phi$.

Assuming that the boundary conditions \eqref{BCH} are satisfied simultaneously, 
both $\mc{H}^q$ and $\mc{H}^p$ will be maximized by the Navier-Stokes system. 
However, since they do not contain derivatives of the Clebsch potentials, 
they are expected to dissipate at a slower rate than functionals such as
the fluid energy or enstrophy. 
Hence, in analogy with Taylor relaxation  
in the context of magnetohydrodynamics where magentic energy is dissipated by
resistivity while the magnetic helicity is kept constant \cite{Taylor1,Taylor2}, one may
conjecture that there exists a time scale where the entropy measures 
$\mc{H}^q$ and $\mc{H}^p$ remain approximately constant, while kinetic energy $E=\frac{1}{2}\int_{\Omega}\bol{v}^2\,dV$ 
approaches a minimum value due to the effect of kinematic viscosity. 
The resulting `transient' steady state is obtained through the 
variational problem
\begin{equation} 
\delta\left[E-\beta^{-1}\lr{\mc{H}^q+\mc{H}^p}\right]=0,\label{Var}
\end{equation}
where variations are carried out with respect to the Clebsch parameters $\phi$, $p^i$, $q^i$, $i=1,...,N$ and $\beta$ is a spatial constant (Lagrange multiplier). The transient steady state resulting from \eqref{Var} is
\begin{subequations}
\begin{align}
q^i=&-\beta\nabla\phi\cdot\nabla p^i-\beta\sum_{j=1}^Np^j\nabla q^j\cdot\nabla p^i,~~~~i=1,...,N,\\
p^i=&\beta\nabla\phi\cdot\nabla q^i+\beta\sum_{j=1}^Np^j\nabla q^j\cdot\nabla q^i,~~~~i=1,...,N,\\
\Delta\phi=&-\sum_{i=1}^N\nabla\cdot\lr{ p^i\nabla q^i}.
\end{align}
\end{subequations}

\section{Clebsch Parametrization and Vortex Stretching}
Consider the following Clebsch parametrization of the fluid velocity,
\begin{equation}
\bol{v}=\sum_{i=1}^{R}\nabla\mu^i\cp\nabla\lambda^i.\label{Cl2}
\end{equation}
This parametrization applies to incompressible vector fields since the divergence of \eqref{Cl2} is identically zero. 
Furthermore, it is complete whenever $R\geq 2$ and the field $\bol{v}$ is exact, i.e. it can be expressed through a vector potential $\bol{\xi}$ as $\bol{v}=\nabla\cp\bol{\xi}$.  
To see this, recall that given a cotangent basis $\lr{\nabla\alpha,\nabla\beta,\nabla\gamma}$ in $\Omega$, any vector field $\bol{\xi}$
can be expressed as
\begin{equation}
\bol{\xi}=f_{\alpha}\nabla\alpha+f_{\beta}\nabla\beta+f_{\gamma}\nabla\gamma,\label{xi1}
\end{equation}
for some coefficients $f_{\alpha}$, $f_{\beta}$, and $f_{\gamma}$. 
Next, we look for functions $\zeta$, $f_{\alpha}'$ and $f_{\beta}'$ such that
\begin{equation}
\bol{\xi}=\nabla\zeta+f_{\alpha}'\nabla\alpha+f_{\beta}'\nabla\beta.\label{xi2}
\end{equation}
Equating \eqref{xi1} with \eqref{xi2}, one finds the solution
\begin{equation}
\zeta=\int f_{\gamma}d\gamma,~~~~f_{\alpha}'=f_{\alpha}-\int\frac{\p f_{\gamma}}{\p\alpha}d\gamma,~~~~f_{\beta}'=f_{\beta}-\int\frac{\p f_{\gamma}}{\p\beta}d\gamma. 
\end{equation}
Hence, the curl of a vector field $\bol{\xi}$ has general expression
\begin{equation}
\nabla\cp\bol{\xi}=\nabla f_{\alpha}'\times\nabla\alpha+\nabla f_{\beta}'\cp\nabla\beta. 
\end{equation}
It follows that for an incompressible velocity field expressed as $\bol{v}=\nabla\cp\bol{\xi}$, with $\bol{\xi}$ a vector potential, the Clebsch parametrization \eqref{Cl2} is complete whenever $R\geq 2$. 

Now, consider the vortex stretching term \eqref{VS}. 
Using the boundary condition \eqref{BC} and standard vector identities, we have
\begin{equation}
\int_{\Omega}\lr{\bol{v}\cdot\nabla\bol{v}}\cdot\nabla\cp\bol{\omega}\,dV=-\int_{\Omega}\bol{v}\cdot\bol{\omega}\cp\lr{\nabla\cp\bol{\omega}}\,dV.\label{VS2}
\end{equation}
Next, assume that $R=1$, i.e. $\bol{v}=\nabla\mu\cp\nabla\lambda$. 
If $\nabla\mu$ and $\nabla\lambda$ are linearly dependent, 
$\bol{v}=\bol{0}$, and the integrand of the vortex stretching term is identically zero. 
Let $\Pi\subset\Omega$ denote the open set of points where $\nabla\mu$ and $\nabla\lambda$ are linearly independent. 
We construct a third function $\ell$ with the property that
\begin{equation}
\nabla\ell\cdot\nabla\mu\cp\nabla\lambda=\abs{\nabla\mu\cp\nabla\lambda}~~~~{\rm in}~~\Pi.\label{ell}
\end{equation} 
A rigorous way to derive the function $\ell$ consists 
in finding a finite number $k$ of small
neighborhoods $U^1,...,U^k$ that cover the whole $\Pi$, i.e. $\Pi\subset U^1\cup ...\cup U^k$.  
Then, equation \eqref{ell} can be solved in each neighborhood by the method of characteristics, giving $k$ solutions $\ell^1,...,\ell^k$. 
In the intersections $U^i\cap U^j$, $i,j=1,...,k$, the functions
$\ell^i$ and $\ell^j$ differ up to a function $\ell^i-\ell^j=\Delta^{ij}\lr{\mu,\lambda}$ of the variables $\mu$ and $\lambda$, since
$\bol{v}\cdot\nabla\lr{\ell^i-\ell^j}=0$ there. 
One can then define a new solution $\ell$ with domain $U^i\cup U^j$ by first extending the domain of $\Delta^{ij}\lr{\mu,\lambda}$ to $U^j$ (Whitney extension theorem \cite{Whitney}), and then by setting
\begin{equation}
\ell=\ell^i~~~~{\rm in}~~U^i,~~~~\ell=\ell^j+\Delta^{ij}~~~~{\rm in}~~U^j.
\end{equation}
The procedure can be repeated until the domain of the
function $\ell$ covers the whole $\Pi$. 

Then, the cotangent vectors $\nabla\ell$, $\nabla\mu$, and $\nabla\lambda$
can be used to decompose the vector field $\bol{\omega}\cp\lr{\nabla\cp\bol{\omega}}$ in the region $\Pi$ as
\begin{equation}
\bol{\omega}\cp\lr{\nabla\cp\bol{\omega}}=f_{\ell}\nabla\ell+f_{\mu}\nabla\mu+f_{\lambda}\nabla\lambda,
\end{equation}
where $f_{\ell}$, $f_{\mu}$, and $f_{\lambda}$ are the coefficients
of the decomposition. 
It follows that
\begin{equation}
\begin{split}
\bol{v}\cdot\bol{\omega}\cp\lr{\nabla\cp\bol{\omega}}=&f_{\ell}\nabla\mu\cp\nabla\lambda\cdot\nabla\ell\\
=&\bol{v}\cdot\nabla\lr{\int \frac{\bol{\omega}\cp\lr{\nabla\cp\bol{\omega}}\cdot\nabla\mu\cp\nabla\lambda}{\nabla\ell\cdot\nabla\mu\cp\nabla\lambda}d\ell}\\
=&\nabla\cdot\left[\bol{v}\int\lr{\frac{\bol{\omega}\cp\lr{\nabla\cp\bol{\omega}}\cdot\bol{v}}{\abs{\bol{v}}}d\ell}\right].
\end{split}
\end{equation}
Therefore, setting $\Phi=\int \frac{\bol{\omega}\cp\lr{\nabla\cp\bol{\omega}}\cdot\bol{v}}{\abs{\bol{v}}}d\ell$, the vortex stretching integral is
\begin{equation}
\int_{\Omega}\lr{\bol{v}\cdot\nabla\bol{v}}\cdot\nabla\cp\bol{\omega}\,dV=-\int_{\Pi}\bol{v}\cdot\bol{\omega}\cp\lr{\nabla\cp\bol{\omega}}\,dV=-\int_{\Pi}\nabla\cdot\lr{\Phi\bol{v}}\,dV=-\int_{\p\Pi}\Phi\bol{v}\cdot\bol{n}'\,dS=0.
\end{equation}
Here, $\p\Pi$ denotes the boundary of $\Pi$, $\bol{n}'$ the unit outward normal to $\p\Pi$, and we used the
fact that by construction $\bol{v}=\bol{0}$ on $\p\Pi$, and the domain of the function $\Phi$ can be extended to include $\p\Pi$. 
 
We conclude by observing that for a two dimensional flow $\bol{v}=\nabla\psi\cp\nabla z$ one has
\begin{equation}
\bol{v}\cdot\bol{\omega}\cp\lr{\nabla\cp\bol{\omega}}=-\Delta\psi\nabla\psi\cp\nabla z\cdot\nabla z\cp\lr{\nabla\Delta\psi\cp\nabla z}=-\frac{1}{2}\nabla\cdot\left[\bol{v}\lr{\Delta\psi}^2\right].
\end{equation}
Hence, the vortex stretching integral vanishes. 

\section{Concluding Remarks}

In this paper, we studied the incompressible Navier-Stokes equations by using a Clebsch parametrization of the velocity field. 
The Clebsch parametrization is complete, i.e. an arbitrary vector field in three dimensional
Euclidean space can be expressed by appropriate choice of the Clebsch potentials. 
Each Clebsch potential obeys a transport equation with a diffusion operator
that differs, in general, from the standard Laplacian. The discrepancy 
is expressed by a term involving the Lie bracket of the corresponding Clebsch pair. 
The transport equations for the Clebsch potentials define a map sending an arbitrary
initial condition $\bol{v}_0$ to a corresponding solution of the Navier-Stokes system \eqref{NS}, \eqref{BC} 
at time $t$. 

The Clebsch potentials $q^i$, $i=1,...,N$, can be used to define a Shannon-type entropy measure ${\rm H}$, 
i.e. a functional whose growth rate is non-negative.
The entropy measure ${\rm H}$ quantifies the topological complexity of the velocity field $\bol{v}$, 
and it is mathematically different from the kinetic energy because  
it does not contain derivatives of the Clebsch potentials.  
Due to the H theorem, the maximum of ${\rm H}$ is achieved at equilibrium. 
The entropy maximum is characterized by a complete loss of vorticity, 
which implies vanishing of the velocity field as a consequence of boundary conditions.
 
We then introduced a second type of Clebsch parametrization 
and showed that the vortex stretching term appearing in the budget
equation for the fluid entrophy identically vanishes
whenever the velocity field admits the representation
$\bol{v}=\nabla\mu\cp\nabla\lambda$, which is a generalization of the case of 
two dimensional flows $\bol{v}=\nabla\psi\cp\nabla z$.

\section*{Acknowledgment}
The research of NS was partially supported by JSPS KAKENHI Grant No. 17H01177.
The author is grateful to Professor Z. Yoshida for useful discussion on the analysis of the Navier-Stokes system
and the Clebsch representation of vector fields.

\section*{Data Availability}

The data that support the findings of this study are available from
the corresponding author upon reasonable request.

\end{document}